\documentclass[a4paper]{VisionStyle}
\usepackage{epsfig}

\newcommand{\gtae}{$\buildrel {\lower3pt\hbox{$>$}} \over 
{\lower2pt\hbox{$\sim$}} $}
\newcommand{\ltae}{$\buildrel {\lower3pt\hbox{$<$}} \over
{\lower2pt\hbox{$\sim$}} $}

\begin{document}

\title{Discrete and diffuse X-ray emission in the nucleus and disk 
of the starburst spiral galaxy M83}

\author{R.\,Soria\inst{1} \and K.\,Wu\inst{1}} 

\institute{Mullard Space Science Laboratory, 
          University College London, Holmbury St Mary, 
          Surrey RH5 6NT, UK }

\maketitle 

\begin{abstract}

We have studied the face-on, barred spiral M83 (NGC 5236) with {\em Chandra}.
       Eighty-one point sources are detected (above 3.5-$\sigma$) 
        in the ACIS S3 image:  
        15 of them are within the inner 16\arcsec~region 
(starburst nucleus, resolved for the first time 
with {\em Chandra}), and 23 within the inner 60\arcsec~(including the bar).
          The luminosity distributions of the sources 
in the inner 60\arcsec~region (nucleus and stellar bar) is a single power law, 
which we interpret as due to continuous, ongoing star formation. 
Outside this inner region, there is 
a smaller fraction of bright sources, which we interpret 
as evidence of an aging population from a past episode of star formation.
      About 50\% of the total emission in the nuclear region  
        is unresolved; of this, about 70\% can be attributed 
	to hot thermal plasma, and we interpret the rest 
as due to unresolved point sources (eg, faint X-ray binaries). 
The unresolved X-ray emission also shows differences 
between the nuclear region and the spiral arms. In the nuclear region, 
the electron temperature of the thermal plasma is $\approx 0.58$ keV. In the 
spiral arms, the thermal component is at $kT \approx 0.31$ keV 
and a power-law component dominates at energies \gtae~1 keV. 
The high abundance of C, Ne, Mg, Si and S with respect to Fe 
suggests that the interstellar medium is enriched and heated  
by core-collapse supernova explosions and winds from massive stars.

\end{abstract}

\section{Introduction}
  
M83 (NGC~5236) is a grand-design, barred spiral galaxy 
   (Hubble type SAB(s)c) with a starburst nucleus.  
Distance estimates are still very uncertain.
A value of 3.7~Mpc was obtained 
   by \cite*{rsoria-E3:va91}.  
This places the galaxy in the Centaurus A group,  
  whose members have a large spread in morphology and high velocities,  
  indicating that the group is not virialised 
  and tidal interactions and merging are frequent 
  (\cite{rsoria-E3:va79}; \cite{rsoria-E3:co97}).  

M83 was observed in the X-ray bands by {\em Einstein} 
  in 1979--1981 (\cite{rsoria-E3:tr85}), 
  by {\em ROSAT} in 1992--1994 (\cite{rsoria-E3:im99}), and 
  by {\em ASCA} in 1994 (\cite{rsoria-E3:ok97}).   
Twenty-one point sources were found in the {\em ROSAT}/HRI image, 
but the starburst nuclear region was unresolved.    

M83 was observed by {\em Chandra} on 2000 April 29, 
  with the ACIS-S3 chip at the focus.
The data became available to the public in mid-2001. 
The total exposure time was 50.978~ks; after screening out 
observational intervals corresponding to background 
flares, we retained a good time interval of 49.497~ks. 
    
In this paper we present the luminosity distribution of 
the discrete source population and discuss the properties 
of the unresolved emission in the nuclear region and in the disk.   
For further details on the data analysis techniques, 
and for more extensive discussions 
on the properties of the individual sources, see \cite*{rsoria-E3:sw02}.

\section{Global properties of the discrete sources}    
\label{rsoria-E3_sec:glob}

A total of 81 point sources 
  are detected in the S3 chip at a 3.5-$\sigma$ level 
in the 0.3--8.0~keV band. The source list is given in 
\cite*{rsoria-E3:sw02}.    
Comparing the position of the {\em Chandra} S3 sources 
with a VLT $B$ image  
  shows that the off-centre sources tend to associate 
  with the optically bright regions (Figure 1).  
The sources have a large spread in the hardness of their X-ray emission.  
  A ``true-colour'' X-ray image of the nuclear region is shown 
  in Figure 2, bottom panel. 

Separating the sources inside and outside a circular region 
  of radius 60\arcsec~from the geometric centre of the X-ray emission   
  reveals that the two groups have different luminosity distributions 
in the 0.3--8.0~keV band.
(A linear separation of 60\arcsec~corresponds to 1.1~kpc 
  for a distance of 3.7~Mpc, 
  and is roughly half of the total length of the major galactic bar.)
The cumulative log~N($>$S) -- log~S distribution 
(where S are the photon counts)
of the sources inside this inner region  
  can be described as a single power law, with a slope of $-0.8$.
The log~N($>$S) -- log~S curve 
  of the sources outside the circular inner region instead  
  is neither a single nor a broken power law (Figure 3).       
It shows a kink at S $\approx 250$~cts;
the slope of the curve above the kink is $-1.3$, while 
it is $-0.6$ at the faint end.
If we assume a foreground absorbing column density 
  $n_{\rm H} = 4 \times 10^{20}$ cm$^{-2}$
  (\cite{rsoria-E3:sc98}), 
a distance of 3.7~Mpc and  
a power-law spectrum with photon index $\Gamma = 1.5$ 
for all the sources, 
  100~counts ($\approx 2.0 \times 10^{-3}$~cts~s$^{-1}$) 
  correspond to an unabsorbed source luminosity 
  $L_{\rm x} = 2.3 \times 10^{37}$~erg~s$^{-1}$ in the 0.3--8.0~keV band.  
The kink in the log~N($>$S) -- log~S curve 
  of the sources outside the 60\arcsec~circle 
  is therefore located at 
$L_{\rm x} \approx 6 \times 10^{37}$~erg~s$^{-1}$ (0.3--8.0~keV band).   
    
We estimate from the Deep Field South survey (\cite{rsoria-E3:gi01})
that about 15\% of the 81 sources are background AGN; 
the expected number in the inner 60\arcsec~circle  
   is smaller than one. The kink in the log~N($>$S) -- log~S curve 
   for the outer sources and the values of the slope at both ends 
are unaffected by the background subtraction.

\begin{figure}[t]
  \begin{center}
    \epsfig{file=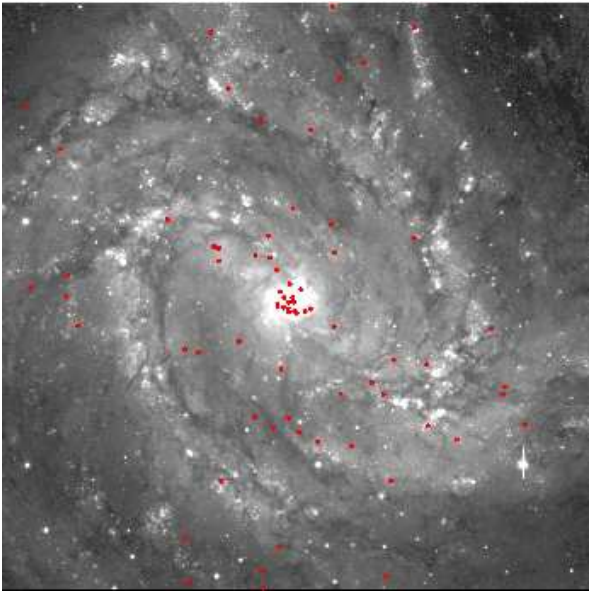, width=7.8cm}
  \end{center}
\caption{Spatial distribution of the discrete X-ray sources overplotted 
      on a VLT $B$ image 
      (size: $6\farcm8 \times 6\farcm8$). }  
\label{rsoria-E3_fig:fig1}
\end{figure}

\begin{figure}[ht]
  \begin{center}
    \epsfig{file=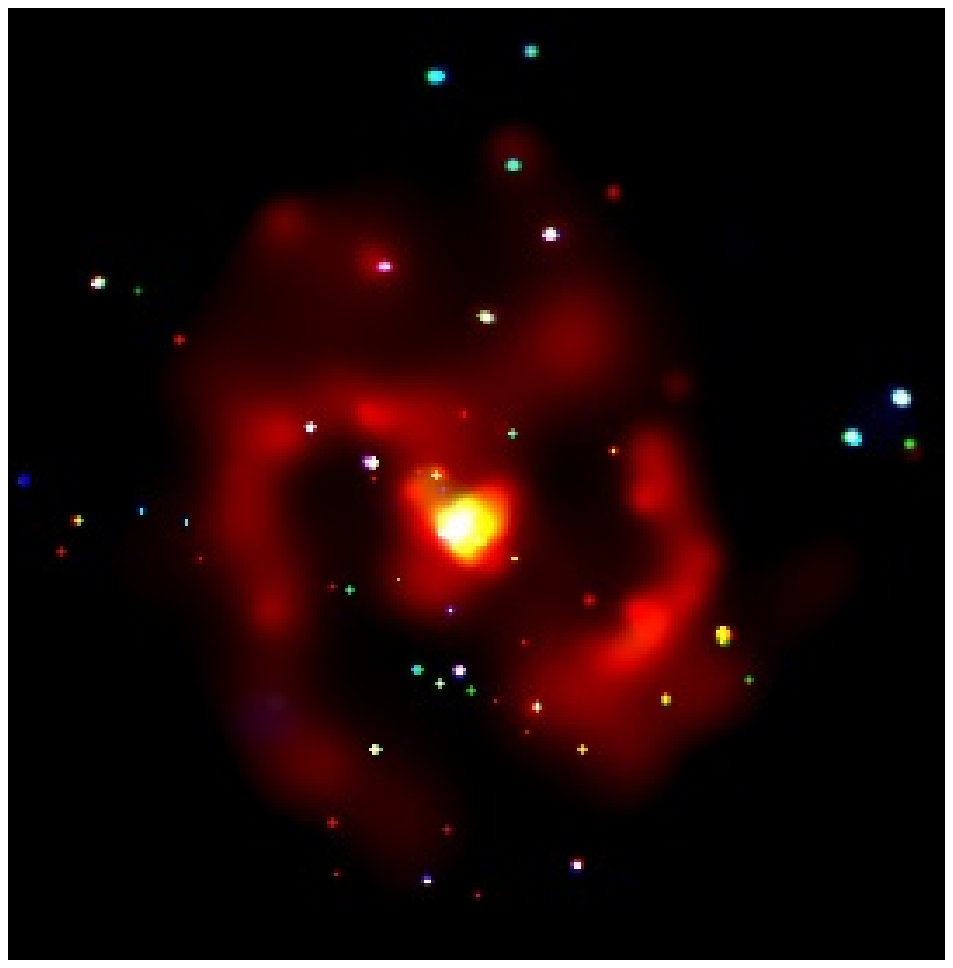, width=7.6cm}\\
    \epsfig{file=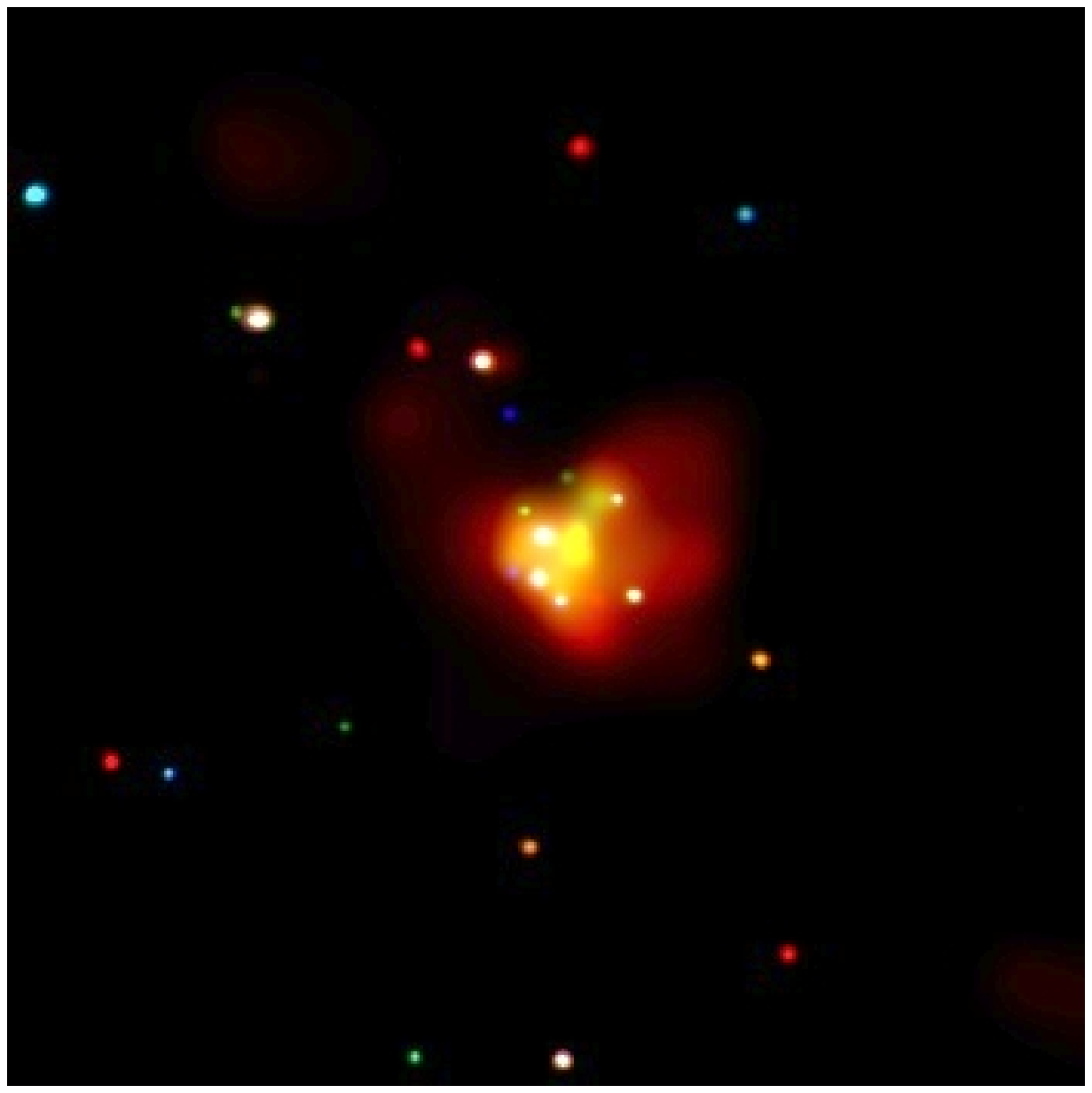, width=7.6cm}
  \end{center}
\caption{ 
Top panel: A low-resolution (ACIS-S3 chip, 
with pixels binned $4 \times 4$) true-colour 
image of M83 shows strong unresolved emission both 
in the starburst nuclear region and along the spiral arms.
Bottom panel: full-resolution true-colour 
image of the central region (size: $3\arcmin \times 3\arcmin$).
In both images, 
red, green and blue correspond to emission in the 0.3--1.0~keV, 
      1.0--2.0~keV and 2.0--8.0~keV bands respectively.
Both images have been adaptively smoothed.} 
\label{rsoria-E3_fig:fig2}
\end{figure}

\begin{figure}[ht]
  \begin{center}
    \epsfig{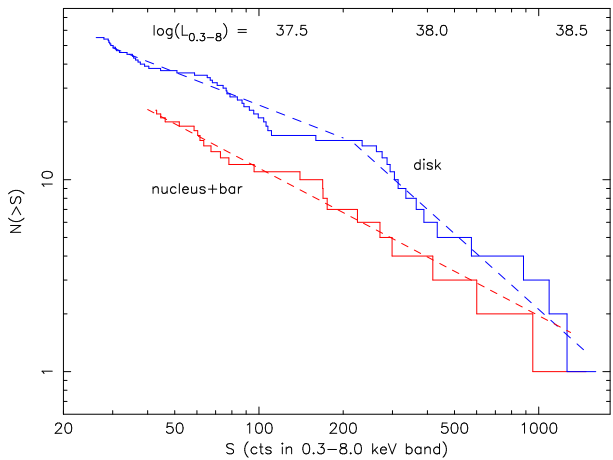}
  \end{center}
\caption{The cumulative luminosity distributions of sources found 
in the 0.3--8.0~keV band is different for the population  
inside the 60\arcsec~inner circle (which includes the starburst 
nuclear region and the main bar) and for the population outside of it (red 
and blue curves respectively). We interpret the single power-law 
distribution of the nuclear sources as evidence of continuous, ongoing 
star formation. The relative scarcity of bright sources in the disk 
population may be the effect of aging, which suggests 
that star formation is less active there. 
See Section 2 for the conversion from counts to luminosity.}  
\label{rsoria-E3_fig:fig3}
\end{figure}

The flatter slope of the log~N($>$S) -- log~S curve 
at the high-luminosity end for the population of sources 
in the inner 60\arcsec~region
 implies a larger proportion of bright sources than 
in source population further away from the nucleus.
The situation is different for example 
  in the spiral galaxy M81, 
  where most bright sources are found in the galactic disk 
  instead of the nuclear region (\cite{rsoria-E3:te01}).  
If the flatness of the slope in the log~N($>$S) -- log~S curve 
  is a characteristic of ongoing star formation (\cite{rsoria-E3:wu01}), 
  the difference in the spatial distribution of the brightest sources 
  in M83 and M81 is simply a consequence of the fact that 
  M83 has a starburst nucleus  
  while star formation in galaxies such as M81 is presently more efficient 
  in the disk. 

\section{Emission from the nuclear region}
    
The {\em Chandra} data reveal that 
  M83 has a very highly structured nuclear region 
  (Figure~2, bottom panel).  
Fifteen discrete sources are detected  
  within a radius of $\approx 16\arcsec$~($\approx 290$ pc)
  from the centre of symmetry of the outer optical isophotes.
A spectral analysis of the two brightest sources is presented 
in \cite*{rsoria-E3:sw02}, and an analysis of other point sources 
will be presented in Soria et~al.~(in preparation).
We removed the point sources and extracted counts 
  from concentric annuli to construct radial brightness profiles  
  of the unresolved emission in the nuclear region.  
We found that the brightness is approximately constant  
  in a circular region up to a radius of 7\arcsec~and 
  then declines radially with a power-law like profile. 
The azimuthally-averaged profile is well fitted by a King profile, 
  with a core radius of $6\farcs7 \pm 0\farcs5$ ($\approx 120$~pc), 
  and a power law with a slope of $-1.9 \pm 0.2$ 
  beyond the core.

We extracted the spectrum of the unresolved emission inside 
the inner 16\arcsec~circle, excluding the resolved point sources, 
and we fitted it using an absorbed, single-temperature 
vmekal plus power-law model.  
Assuming solar abundances, we obtain 
a best-fit electron temperature $kT = (0.60^{+0.02}_{-0.03})$~keV 
and a power-law photon index $\Gamma = 3.1^{+0.1}_{-0.2}$. 
The predicted lines are not strong enough  
  to account for the data, leading to poor fit statistics 
($\chi^2_{\nu} = 1.42$, 114~dof).
Increasing the abundance of all the metals by the same constant 
factor does not improve the fit. 

We then assumed a different set of abundances, higher 
than solar for C, Ne, Mg, Si and S, and 
slightly underubandant for Fe (see Table 1 in \cite{rsoria-E3:sw02}). 
This is physically justified if the interstellar medium 
has been enriched by type-II supernova ejecta and winds from 
very massive, young stars. We obtain a best-fit   
temperature $kT = (0.58^{+0.03}_{-0.02})$~keV, and power-law 
photon index $\Gamma = 2.7^{+0.3}_{-0.3}$   
($\chi^2_{\nu} = 0.99$, 114~dof). 

\begin{table}
  \caption{Luminosity of emission from 
   discrete and unresolved sources in the nuclear region (0.3--8.0 keV band).}
  \label{rsoria-E3_tab:tab1}
  \begin{center}
    \leavevmode
    \footnotesize
    \begin{tabular}[h]{@{}lrr}
\hline \\ [-5pt]
  luminosity ($\times 10^{38}$~erg~s$^{-1}$) & inside 7\arcsec & inside 16\arcsec \\ [5pt] 
\hline \\[-5pt]
discrete sources  & $7.7$ & $12.3$\\[5pt]
unresolved sources  & $8.0$ & $11.5$\\[5pt]
{\hspace*{0.3cm}}optically-thin thermal component   & $5.5$ & $7.8$\\[5pt] 
{\hspace*{0.3cm}}power-law  component    & $2.5$ & $3.7$\\[5pt] 
\hline \\ 
total      & $15.7$ & $23.8$\\[5pt]
\hline \\ 
      \end{tabular}
  \end{center}
\end{table}

We also estimated the total (resolved plus unresolved) 
luminosity from the circular regions within radii 
of 7\arcsec~(this is approximately the region inside the outer dust ring) 
and 16\arcsec~from the geometric centre of the X-ray emission, 
using an absorbed, optically-thin thermal plasma  
plus power-law model. The total emitted luminosity 
in the 0.3--8.0 keV band is $\approx 15.7 \times 10^{38}$~erg~s$^{-1}$ 
inside 7\arcsec~and $\approx 23.8 \times 10^{38}$~erg~s$^{-1}$ 
inside 16\arcsec~(Table 1). Discrete sources contribute 
$\approx 50\%$ of the total luminosity.
The unresolved emission is itself the sum of truly diffuse emission 
from optically thin gas, and emission from unresolved point-like 
sources (eg, faint X-ray binaries). Assuming that the latter 
contribution is responsible for the power-law component 
in the spectrum of the unresolved emission, we estimate that 
emission from truly diffuse thermal plasma contributes $\approx 35\%$ 
of the total luminosity (Table 1).

Extrapolating the log~N($>$S) -- log~S curve for the nuclear sources 
gives us another way of estimating 
the relative contribution to the 
unresolved emission of truly diffuse gas and faint point-like   
sources. We obtain that 
unresolved point-like X-ray sources inside 16\arcsec~would 
have a total luminosity 
  of $\approx 2.7 \times 10^{38}$ erg s$^{-1}$ (see \cite{rsoria-E3:sw02} 
for details).
Another possible contribution to the unresolved emission 
comes from photons emitted by the resolved sources 
but falling outside the extraction regions, in the wings 
of the PSF. We estimate that this contribution is  \ltae~$1.5 \times 
10^{38}$~erg~s$^{-1}$. 
Thus, the combined contribution of faint X-ray sources and emission in 
the wings of the PSF can account for 
the luminosity of the power-law component inferred 
from the spectral fitting of the unresolved emission. 
This also confirms that a substantial proportion ($\approx 70$\%) 
of the unresolved emission is indeed due to truly diffuse gas 
rather than faint point-like sources.

\begin{figure}
  \begin{center}
    \epsfig{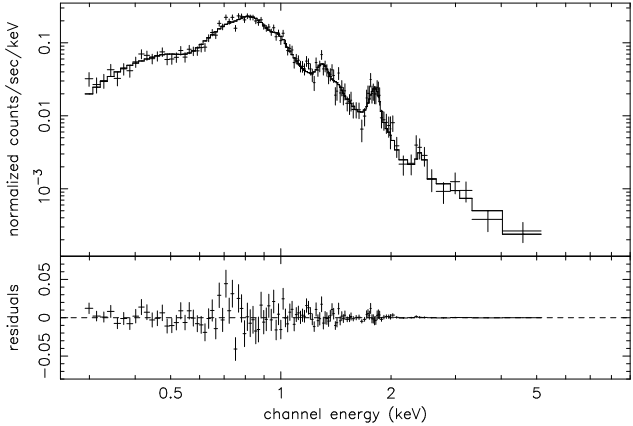}
  \end{center}
\caption{ 
The spectrum of the unresolved emission in the nuclear region (inside a 16\arcsec~circle)   
  shows very strong emission lines from 
  Mg\,{\small{XI}} (1.33--1.35~keV), Si\,{\small{XIII}} (1.84--1.87~keV) and S\,{\small{XV}} (2.43 keV).
The other major contributions to the spectrum come from 
C\,{\small{VI}} (0.37--0.44~keV), 
O\,{\small{VII}} (triplet, 0.56--0.57~keV), 
Ne\,{\small{IX}} (triplet, 0.91--0.92~keV), 
Ne\,{\small{X}} (1.02~keV), 
and the Fe\,{\small{XVII}} L-line complex (0.6--0.9~keV).}  
\label{rsoria-E3_fig:fig4}
\end{figure}

\begin{figure}
  \begin{center}
    \epsfig{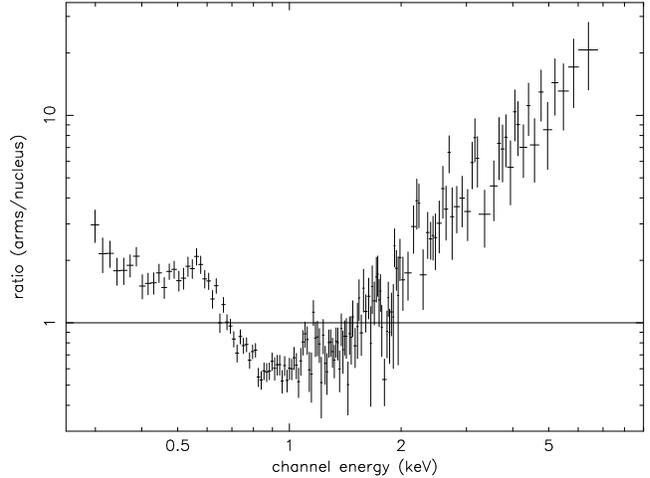}
  \end{center}
\caption{The ratio between the unresolved 
emission in the spiral arms and the best-fit model to 
the unresolved emission in the nuclear region shows that: 
a) the thermal component in the spiral arms is cooler 
($T \approx 0.31$ keV in the arms, $T \approx 0.58$ keV in the nucleus); 
b) a power-law component dominates the spiral-arm emission 
at energies \gtae~1 keV.
The two spectra have 
been normalised to the same number of counts. }  
\label{rsoria-E3_fig:fig5}
\end{figure}

Diffuse X-ray emission is also clearly observed 
along the spiral arms (Figure~2, top panel). 
We compared the spectrum of the unresolved emission in the nuclear region 
and along the arms, and found that the unresolved arm 
emission has a thermal component 
dominating at lower energies, and a power-law-like component, 
dominating above 1 keV. The thermal component in the arms has 
a temperature $T = 0.31^{+0.01}_{-0.03}$~keV, much cooler than that 
in the nucleus. The power-law component has 
a photon index $\Gamma = 1.42^{+0.08}_{-0.05}$.
A more detailed analysis of the unresolved arm emission 
will be presented in Soria et~al.~(in preparation).

\section{Discussion}

Separating the discrete sources inside and outside a 60\arcsec~central region  
  reveals that the two populations have different cumulative 
luminosity distributions.    
The log~N($>$S) -- log S curve of the sources outside this radius 
(i.e., the disk population) shows a kink at a luminosity 
$\approx 6 \times 10^{37}$~erg~s$^{-1}$ in the 0.3--8.0 keV band. 
No kink is seen for the sources inside the 60\arcsec~radius 
(i.e., those located in the nuclear region and along the bar). 
The slope of the log~N($>$S) -- log S curve at its high-luminosity end 
is flatter for the nuclear population, implying a larger proportion 
of bright sources (possible black-hole candidates). We interpret 
this as evidence of a past star formation episode in the disk, 
while there is continuous, ongoing star formation 
in the nuclear region (\cite{rsoria-E3:wu01}).    

About 50\% of the total emission in the nuclear region  
belongs to resolved discrete sources.
We estimate 
that $\approx 70$\% of the unresolved emission 
(35\% of the total) is due to truly diffuse plasma, 
with the rest (15\% of the total) coming from 
faint, unresolved point-like sources and photons in the wings 
of the PSF outside the detection cells of the resolved sources.

The X-ray spectrum of the unresolved nuclear component    
   shows strong emission lines, 
   and can be modelled as emission from optically-thin thermal plasma 
at $kT \approx 0.6$ keV.  Above-solar abundances of Ne, Mg, Si and S 
are required to fit the spectrum, while Fe appears to be underabundant. 
This suggests that the interstellar medium in the starburst nuclear 
region has been enriched by the ejecta of type-II 
supernova explosions. Moreover, 
a high abundance of C and a high C/O abundance ratio  
can be the effect of radiatively-driven winds from metal-rich Wolf-Rayet stars 
with $M$ \gtae~$40$\,M$_{\odot}$ 
(\cite{rsoria-E3:gu99}). Both effects are likely to be present 
in the nuclear region.  

Strong unresolved emission is also detected along the arms. 
It is well fitted by a thermal component at $kT \approx 0.3$ keV 
and a power-law component ($\Gamma \approx 1.5$) 
dominating at energies \gtae~1 keV. A study of 
this higher-energy component and a comparison with the Galactic Ridge 
emission will be presented in a work now in preparation.

\begin{acknowledgements}

     We thank Stefan Immler, Roy Kilgard, 
Miriam Krauss, Casey Law, Oak-Young Park, Elena Pian, 
Allyn Tennant and Daniel Wang for helpful discussions and suggestions.

\end{acknowledgements}


\begin{thebibliography}{}


\bibitem[\protect\astroncite{C{\^ o}t{\' e} et~al.}{1998}]{rsoria-E3:co97} 
      C{\^ o}t{\' e}, S., Freeman, K. C., Carignan, C. \& Quinn, P. 
      1997, AJ, 114, 1313 

\bibitem[\protect\astroncite{Giacconi et~al.}{2001}]{rsoria-E3:gi01} 
      Giacconi, R., et al. 
      2001, ApJ, 551, 624 

\bibitem[\protect\astroncite{Gustafsson et~al.}{1999}]{rsoria-E3:gu99}
      Gustafsson, B., Karlsson, T., Olsson, E., Edvardsson, B. \& Ryde, N. 
      1999, A\&A, 342, 426

\bibitem[\protect\astroncite{Immler et~al.}{1999}]{rsoria-E3:im99} 
      Immler, S., Volger, A., Ehle, M. \& Pietsch, W. 
      1999, A\&A, 352, 415    

\bibitem[\protect\astroncite{Okada et~al.}{1997}]{rsoria-E3:ok97} 
      Okada, K., Mitsuda, K. \& Dotani, T. 
      1997, PASJ, 49, 653 


\bibitem[\protect\astroncite{Sandage \& Tamman}{1987}]{rsoria-E3:sa87}   
      Sandage, A. \& Tammann, G. A. 
      1987, A revised Shapley-Ames Catalog of Bright Galaxies, 2nd ed.,  
      (Carnegie Institution of Washington Publication: Washington)

\bibitem[\protect\astroncite{Schlegel et~al.}{1998}]{rsoria-E3:sc98}
      Schlegel, D. J., Finkbeiner, D. P. \& Davis, M.
      1998, ApJ, 500, 525

\bibitem[\protect\astroncite{Soria \& Wu}{2002}]{rsoria-E3:sw02}
Soria, R., Wu, K., 2002, A\&A, in press (astro-ph/0201059)
 

\bibitem[\protect\astroncite{Tennant et~al.}{2001}]{rsoria-E3:te01} 
   Tennant, A. F., Wu, K., Ghosh, K. K., Kolodziejczak, J. J. \& Swartz, D. A. 
      2001, ApJ, 549, L43

\bibitem[\protect\astroncite{Trinchieri et~al.}{1985}]{rsoria-E3:tr85} 
      Trinchieri, G., Fabbiano, G. \& Palumbo, G. G. C.
      1985, ApJ, 290, 96

\bibitem[\protect\astroncite{de Vaucouleurs}{1979}]{rsoria-E3:va79}   
      de Vaucouleurs, G. 
      1979, AJ, 84, 1270 

\bibitem[\protect\astroncite{de Vaucouleurs et~al.}{1991}]{rsoria-E3:va91} 
      de Vaucouleurs, G., de Vaucouleurs, A., Corwin, H., Jr., Buta, R., 
      Paturel, G. \& Fouque, P. 
      1991, Third Reference Catalogue of Bright Galaxies (Springer: New York)  

  
\bibitem[\protect\astroncite{Wu}{2001}]{rsoria-E3:wu01} 
      Wu, K. 
      2001, PASA, 18, 443 

\end{thebibliography}
\end{document}